\documentclass[aps,preprint,amsfonts,amssymb,amsmath,%
tightenlines,nofootinbib,showpacs,showkeys]{revtex4}

\newcommand{\cN}{\mathcal{N}}
\newcommand{\cP}{\mathcal{P}}
\newcommand{\cT}{\mathcal{T}}
\newcommand{\cV}{\mathcal{V}}
\newcommand{\bH}{\boldsymbol{H}}
\newcommand{\bQ}{\boldsymbol{Q}}
\newcommand{\fF}{\mathfrak{F}}
\newcommand{\bbN}{\mathbb{N}}

\newcommand{\rmd}{\mathrm{d}}

\newcommand{\sPN}{\mathsf{P}_{\!\cN}}
\newcommand{\sV}{\mathsf{V}}

\begin{document}



%
%

\title{$\cN$-fold Parasupersymmetry}
\author{Toshiaki Tanaka}
\email{ttanaka@mail.ncku.edu.tw}
\affiliation{Department of Physics, National Cheng-Kung University,\\
 Tainan 701, Taiwan, R.O.C.\\
 National Center for Theoretical Sciences, Taiwan, R.O.C.}


\begin{abstract}

We find a new type of non-linear supersymmetries, called $\cN$-fold
parasupersymmetry, which is a generalization of both $\cN$-fold
supersymmetry and parasupersymmetry. We provide a general formulation
of this new symmetry and then construct a second-order $\cN$-fold
parasupersymmetric quantum system where all the components of
$\cN$-fold parasupercharges are given by type A $\cN$-fold supercharges.
We show that this system exactly reduces to the Rubakov--Spiridonov
model when $\cN=1$ and admits a generalized type C $2\cN$-fold
superalgebra. We conjecture the existence of other `$\cN$-fold
generalizations' such as $\cN$-fold fractional supersymmetry,
$\cN$-fold orthosupersymmetry, and so on.
\end{abstract}






\maketitle

\section{Introduction}

It goes without saying that the concept of symmetry has played a central
role in the development of modern theoretical physics and mathematical
science. This means in particular that a new discovery of a new symmetry
enlarges our ability and possibility to describe new phenomena in any
field of science, regardless of whether or not they have already been
observed in nature.
In this letter, we report on a new type of non-linear supersymmetries,
called $\cN$-fold parasupersymmetry, which is a generalization of both
$\cN$-fold supersymmetry \cite{AIS93,AST01b,AS03} and parasupersymmetry
\cite{RS88,BD90a,To92,Kh92,CF03}. In our previous paper \cite{Ta07a},
we generically formulated parasupersymmetry and showed that
parasupersymmetric quantum systems as well as $\cN$-fold supersymmetric
ones are weakly quasi-solvable. The latter fact implies that corresponding
higher-dimensional quantum field theories, if exist, would satisfy some
kinds of perturbative non-renormalization theorems, as in the case of
ordinary supersymmetric quantum field theory, since quasi-solvability is
a one-dimensional analog and, in a sense, a generalization of the
theorems~\cite{AST01b}. Hence, $\cN$-fold parasupersymmetric quantum
systems presented in this letter would provide up to now the most
general framework for describing
a physical system which has characteristic features like some
non-renormalization theorems. For the formulation, we fully
employ the general formalism of parafermionic algebra and
parasupersymmetry previously proposed by us in Ref.~\cite{Ta07a}
and omit technical details of them in this letter. Hence, for
the details see Ref.~\cite{Ta07a} and the references cited therein.

\section{Definition}

First of all, let us introduce parafermionic algebra of order
$p(\in\bbN)$.
It is an associative algebra composed of the identity operator $I$
and two parafermionic operators $\psi^{-}$ and $\psi^{+}$ of order
$p$ which satisfy the nilpotency:
\begin{align}
(\psi^{-})^{p}\neq 0,\quad (\psi^{+})^{p}\neq 0,\qquad
 (\psi^{-})^{p+1}=(\psi^{+})^{p+1}=0.
\label{eq:nilpo}
\end{align}
Hence, we immediately have $2p+1$ non-zero elements
$\{I,\psi^{-},\dots,(\psi^{-})^{p},\psi^{+},\dots,(\psi^{+})^{p}\}$.
We call them the \emph{fundamental} elements of parafermionic
algebra of order $p$. Parafermionic algebra is characterized by
anti-commutation relation $\{A,B\}=AB+BA$ and commutation relation
$[A,B]=AB-BA$ among the fundamental elements.

We shall next define parafermionic Fock spaces $\sV_{p}$ of order $p$
on which the parafermionic operators act. The latter space is $(p+1)$
dimensional and its $p+1$ bases $|k\rangle$ ($k=0,\dots,p$)
are defined by
\begin{align}
\psi^{-}|0\rangle=0,\quad |k\rangle=(\psi^{+})^{k}|0\rangle,
 \quad\psi^{-}|k\rangle=|k-1\rangle\quad (k=1,\dots,p).
\label{eq:defpfs}
\end{align}
That is, $\psi^{-}$ and $\psi^{+}$ act as annihilation and
creation operators of parafermions, respectively.
The state $|0\rangle$ is called the parafermionic \emph{vacuum}.
The subspace spanned by each state $|k\rangle$ ($k=0,\dots,p$) is
called the $k$-parafermionic subspace and is denoted by
$\sV_{p}^{(k)}$.
We can now define a set of projection operators
$\Pi_{k}:\sV_{p} \to\sV_{p}^{(k)}$ ($k=0,\dots,p$) which satisfy
\begin{align}
\Pi_{k}|l\rangle=\delta_{k,l}|k\rangle,\qquad\Pi_{k}\Pi_{l}=
 \delta_{k,l}\Pi_{k},\qquad\sum_{k=0}^{p}\Pi_{k}=I.
\label{eq:defpo}
\end{align}
From the definitions (\ref{eq:defpfs}) and (\ref{eq:defpo}),
we obtain
\begin{align}
\Pi_{k+1}\psi^{+}=\psi^{+}\Pi_{k},\qquad
 \psi^{-}\Pi_{k+1}=\Pi_{k}\psi^{-},
\label{eq:Pipsi}
\end{align}
where and hereafter we put $\Pi_{k}\equiv 0$ for all $k<0$ and
$k>p$.\\

Parasupersymmetry of order 2 in quantum mechanics was first introduced
by Rubakov and Spiridonov \cite{RS88} and was later generalized
to arbitrary order independently by Tomiya~\cite{To92} and by
Khare~\cite{Kh92}. A different formulation for order 2 was proposed
by Beckers and Debergh \cite{BD90a} and a generalization of the latter
to arbitrary order was attempted by Chenaghlou and Fakhri \cite{CF03}.
Thus, we call them RSTK and BDCF formalism, respectively.
We shall generalize them such that they reduce to $\cN$-fold
supersymmetry in Ref.~\cite{AST01b} when parafermionic order is 1.
For this purpose, we first introduce a pair of $\cN$-fold
parasupercharges $\bQ_{\cN}^{\pm}$ of order $p$ which satisfy
\begin{align}
(\bQ_{\cN}^{-})^{p}\neq 0,\quad(\bQ_{\cN}^{+})^{p}\neq 0,\quad
 (\bQ_{\cN}^{-})^{p+1}=(\bQ_{\cN}^{+})^{p+1}=0.
\label{eq:pfsc1}
\end{align}
A system $\bH$ is said to have \emph{$\cN$-fold parasupersymmetry of
order $p$} if it commutes with the $\cN$-fold parasupercharges of
order $p$
\begin{align}
[\bQ_{\cN}^{-},\bH]=[\bQ_{\cN}^{+},\bH]=0,
\label{eq:pfsc2}
\end{align}
and satisfies the non-linear relations in a generalized RSTK formalism
\begin{subequations}
\label{eq:pfsc3}
\begin{align}
\sum_{k=0}^{p}(\bQ_{\cN}^{-})^{p-k}\bQ_{\cN}^{+}(\bQ_{\cN}^{-})^{k}
 =C_{p}(\bQ_{\cN}^{-})^{p-1}\sPN (\bH),\\
\sum_{k=0}^{p}(\bQ_{\cN}^{+})^{p-k}\bQ_{\cN}^{-}(\bQ_{\cN}^{+})^{k}
 =C_{p}\sPN (\bH)(\bQ_{\cN}^{+})^{p-1},
\end{align}
\end{subequations}
where $\sPN (x)$ is a monic polynomial of degree $\cN$ in $x$ and
$C_{p}$ is a constant, or in a generalized BDCF formalism
\begin{subequations}
\label{eq:pfsc4}
\begin{align}
\underbrace{[\bQ_{\cN}^{-},\cdots,[\bQ_{\cN}^{-}}_{(p-1)\text{ times}},
 [\bQ_{\cN}^{+},\bQ_{\cN}^{-}]]\cdots]&=(-1)^{p}C_{p}
 (\bQ_{\cN}^{-})^{p-1}\sPN (\bH),\\
\underbrace{[\bQ_{\cN}^{+},\cdots,[\bQ_{\cN}^{+}}_{(p-1)\text{ times}},
 [\bQ_{\cN}^{-},\bQ_{\cN}^{+}]]\cdots]&=C_{p}\sPN (\bH)
 (\bQ_{\cN}^{+})^{p-1},
\end{align}
\end{subequations}
It is evident that in both generalizations (\ref{eq:pfsc3}) and (\ref{eq:pfsc4}),
which are completely new algebraic relations and have never appeared in
the past literature, they reduce to ordinary parasupersymmetry of
the corresponding formulations when $\cN=1$. 
As we pointed out in Ref.~\cite{Ta07a}, an apparent
drawback of the (generalized) BDCF formalism is that the relations
(\ref{eq:pfsc4}) do not reduce to the ordinary $\cN$-fold
supersymmetric anti-commutation relation $\{\bQ^{-},\bQ^{+}\}=C_{1}
\sPN (\bH)$ when $p=1$, in contrast to the RSTK relation
(\ref{eq:pfsc3}). For this reason, we only consider the (generalized)
RSTK formalism in this paper.

An immediate consequence of the commutativity (\ref{eq:pfsc2}) is that
each $n$th-power of the $\cN$-fold parasupercharges ($2\leq n\leq p$)
also commutes with the system $\bH$
\begin{align}
[(\bQ_{\cN}^{-})^{n},\bH]=[(\bQ_{\cN}^{+})^{n},\bH]=0\quad
 (2\leq n\leq p).
\label{eq:pfsc5}
\end{align}
Hence, every $\cN$-fold parasupersymmetric system $\bH$ satisfying
(\ref{eq:pfsc2}) always has $2p$ conserved charges.

To realize $\cN$-fold parasupersymmetry in quantum mechanical systems,
we usually consider a vector space $\fF\times\sV_{p}$ where $\fF$ is
a linear space of complex functions such as the Hilbert space $L^{2}$
in Hermitian quantum theory and the Krein space $L_{\cP}^{2}$ in
$\cP\cT$-symmetric quantum theory \cite{Ta06b,Ta06d}. A parafermionic
quantum system $\bH$ is introduced by
\begin{align}
\bH=\sum_{k=0}^{p}H_{k}\Pi_{k},
\label{eq:pfqs}
\end{align}
where $H_{k}$ ($k=0,\dots,p$) are scalar Hamiltonians acting on $\fF$:
\begin{align}
H_{k}=-\frac{1}{2}\frac{\rmd^{2}}{\rmd q^{2}}+V_{k}(q)\quad
 (k=0,\dots,p).
\label{eq:Schro}
\end{align}
Two $\cN$-fold parasupercharges $\bQ_{\cN}^{\pm}$ are defined by
\begin{align}
\bQ_{\cN}^{-}=\sum_{k=0}^{p}Q_{\cN,k}^{-}\psi^{-}\Pi_{k},
 \qquad\bQ_{\cN}^{+}=\sum_{k=0}^{p}Q_{\cN,k}^{+}\Pi_{k}\psi^{+},
\label{eq:pfsc}
\end{align}
where $Q_{\cN,k}^{+}$ ($k=0,\dots,p$) are $\cN$th-order linear
differential operators acting on $\fF$
\begin{align}
Q_{\cN,k}^{+}=\sum_{l=0}^{\cN}w_{k,l}(q)\frac{\rmd^{l}}{\rmd q^{l}}
 \quad(k=0,\dots,p),
\label{eq:compQ}
\end{align}
and for each $k$ $Q_{\cN,k}^{-}$ is given by a certain `adjoint'
of $Q_{\cN,k}^{+}$, e.g., the (ordinary) adjoint $Q_{\cN,k}^{-}=
(Q_{\cN,k}^{+})^{\dagger}$ in the Hilbert space $L^{2}$,
the $\cP$-adjoint $Q_{\cN,k}^{-}=\cP (Q_{\cN,k}^{+})^{\dagger}\cP$
in the Krein space $L_{\cP}^{2}$, and so on. For all $k\leq 0$ we
put $Q_{\cN,k}^{\pm}\equiv 0$. As we will show shortly, the novel
realization of parafermionic supercharges (\ref{eq:pfsc}) in terms of
$\cN$th-order linear differential operators (\ref{eq:compQ}) indeed
enables us to realize an $\cN$-fold parasupersymmetric system satisfying
(\ref{eq:pfsc2}) and (\ref{eq:pfsc3}). When $p=1$, the triple
$(\bH,\bQ_{\cN}^{-},\bQ_{\cN}^{+})$ defined in Eqs.~(\ref{eq:pfqs})
and (\ref{eq:pfsc}) becomes
\begin{align}
\bH=H_{0}\psi^{-}\psi^{+}+H_{1}\psi^{+}\psi^{-},\quad
\bQ_{\cN}^{-}=Q_{\cN,1}^{-}\psi^{-},\quad
\bQ_{\cN}^{+}=Q_{\cN,1}^{+}\psi^{+},
\end{align}
and thus reduces to an ordinary $\cN$-fold supersymmetric quantum
mechanical system~\cite{AST01b}. The non-linear relation
(\ref{eq:pfsc3}) together with the nilpotency (\ref{eq:pfsc1})
for $p=1$ are just the anti-commutation relations between
supercharges
\begin{align}
\{\bQ_{\cN}^{\pm},\bQ_{\cN}^{\pm}\}=0,\qquad
 \{\bQ_{\cN}^{-},\bQ_{\cN}^{+}\}=C_{1}\sPN (\bH).
\end{align}
Hence, the $\cN$-fold parasupersymmetric quantum systems defined
by Eqs.~(\ref{eq:pfsc1})--(\ref{eq:compQ}) provide a natural
generalization of ordinary $\cN$-fold supersymmetric quantum
mechanics. It is easy to check that the $\cN$-fold parasupercharges
$\bQ^{\pm}$ defined by Eq.~(\ref{eq:pfsc}) already satisfy
the nilpotency (\ref{eq:pfsc1}) and that the commutativity
(\ref{eq:pfsc2}) is satisfied if and only if
\begin{align}
H_{k-1}Q_{\cN,k}^{-}=Q_{\cN,k}^{-}H_{k},\quad 
 Q_{\cN,k}^{+}H_{k-1}=H_{k}Q_{\cN,k}^{+},\quad\forall k=1,\dots,p.
\label{eq:inter}
\end{align}
That is, each pair of $H_{k-1}$ and $H_{k}$ must satisfy the
intertwining relations with respect to the $\cN$th-order linear
differential operators $Q_{\cN,k}^{-}$ and $Q_{\cN,k}^{+}$.
Similarly, the commutativity (\ref{eq:pfsc5}) between
$(\bQ_{\cN}^{\pm})^{n}$ and $\bH$ ($2\leq n\leq p$) means that any
pair of $H_{k-n}$ and $H_{k}$ ($1\leq n\leq k\leq p$) satisfies
\begin{subequations}
\label{eq:Nfold}
\begin{align}
H_{k-n}Q_{\cN,k-n+1}^{-}\cdots Q_{\cN,k-1}^{-}Q_{\cN,k}^{-}
 =Q_{\cN,k-n+1}^{-}\dots Q_{\cN,k-1}^{-}Q_{\cN,k}^{-}H_{k},\\
Q_{\cN,k}^{+}Q_{\cN,k-1}^{+}\dots Q_{\cN,k-n+1}^{+}H_{k-n}
 =H_{k}Q_{\cN,k}^{+}Q_{\cN,k-1}^{+}\cdots Q_{\cN,k-n+1}^{+},
\end{align}
\end{subequations}
which means that $H_{k-n}$ and $H_{k}$ constitute a pair of
$n\cN$-fold supersymmetry. The relations (\ref{eq:Nfold})
can also be derived by repeated applications of Eq.~(\ref{eq:inter}).
Since $\cN$-fold supersymmetry is essentially equivalent to weak
quasi-solvability \cite{AST01b,Ta03a}, $\cN$-fold parasupersymmetric
quantum systems also possess weak quasi-solvability. To see
the structure of weak quasi-solvability in the $\cN$-fold
parasupersymmetric system $\bH$ more precisely, let us first define
\begin{align}
\cV_{n,k}^{-}=\ker (Q_{\cN,k-n+1}^{-}\cdots Q_{\cN,k}^{-}),\quad
 \cV_{n,k}^{+}=\ker (Q_{\cN,k}^{+}\cdots Q_{\cN,k-n+1}^{+})\quad
 (1\leq n\leq k\leq p).
\label{eq:defVnk}
\end{align}
By the definition of (\ref{eq:defVnk}), the vector spaces
$\cV_{n,k}^{\pm}$ for each fixed $k$ are related as
\begin{align}
\cV_{1,k}^{-}\subset\cV_{2,k}^{-}\subset\cdots\subset
 \cV_{k,k}^{-},\quad
\cV_{1,k}^{+}\subset\cV_{2,k}^{+}\subset\cdots\subset
 \cV_{k,k}^{+}.
\label{eq:flag}
\end{align}
On the other hand, it is evident from the intertwining relations
(\ref{eq:Nfold}) that each Hamiltonian $H_{k}$ ($0\leq k\leq p$)
preserves vector spaces as follows:
\begin{subequations}
\label{eq:invsp}
\begin{align}
H_{k}\cV_{n,k}^{-}\subset\cV_{n,k}^{-}\quad(1\leq n\leq k),\\
H_{k}\cV_{n,k+n}^{+}\subset\cV_{n,k+n}^{+}\quad(1\leq n\leq p-k).
\end{align}
\end{subequations}
From Eqs.~(\ref{eq:flag}) and (\ref{eq:invsp}), the largest
space preserved by each $H_{k}$ ($0\leq k\leq p$) is given by
\begin{align}
\cV_{k,k}^{-}+\cV_{p-k,p}^{+}\quad(0\leq k\leq p).
\label{eq:lstinv}
\end{align}
Needless to say, each Hamiltonian $H_{k}$ preserves the two spaces
in Eq.~(\ref{eq:lstinv}) separately.
The intertwining relations (\ref{eq:inter}) and (\ref{eq:Nfold})
ensure that all the component Hamiltonians $H_{k}$ ($k=0,\dots,p$) of
the system $\bH$ are isospectral outside the sectors $\cV_{n,k}^{\pm}$
($1\leq n\leq k\leq p$). The spectral degeneracy of $\bH$ in these
sectors depends on the form of each component of the $\cN$-fold
parasupercharges, $Q_{\cN,k}^{\pm}$ ($k=1,\dots,p$).

In addition to these `power-type' symmetries, every $\cN$-fold
parasupersymmetric quantum system $\bH$ defined in Eq.~(\ref{eq:pfqs})
can have `discrete-type' ones. The conserved charges of this type
are given by
\begin{align}
\bQ_{\cN,\{n\}}^{\pm}=[\{(\psi^{-})^{n},(\psi^{+})^{n}\},\bQ_{\cN}^{\pm}],
 \quad\bQ_{\cN,[n]}^{\pm}=[[(\psi^{-})^{n},(\psi^{+})^{n}],
 \bQ_{\cN}^{\pm}]\quad (n=1,\ldots,p).
\label{eq:discc}
\end{align}
It follows from Jacobi identity that they indeed commute with $\bH$:
\begin{align}
[\bQ_{\cN,\{n\}}^{\pm},\bH]=[\bQ_{\cN,[n]}^{\pm},\bH]=0\quad
 (n=1,\dots,p).
\end{align}
We note, however, that they are in general not linearly independent
and we cannot determine the number of linearly independent conserved
charges without the knowledge of parafermionic algebra of each order.

The non-linear constraints (\ref{eq:pfsc3}) can also be calculated in
a similar way. The first non-linear relation in Eq.~(\ref{eq:pfsc3})
is satisfied if and only if the following two identities hold:
\begin{subequations}
\label{eq:parac}
\begin{align}
Q_{\cN,1}^{-}\cdots Q_{\cN,p}^{-}Q_{\cN,p}^{+}+\sum_{k=1}^{p-1}
 Q_{\cN,1}^{-}\cdots Q_{\cN,p-k}^{-}Q_{\cN,p-k}^{+}Q_{\cN,p-k}^{-}
 \cdots Q_{\cN,p-1}^{-}\notag\\
=C_{p}Q_{\cN,1}^{-}\cdots Q_{\cN,p-1}^{-}\sPN (H_{p-1}),\\
\sum_{k=1}^{p-1}Q_{\cN,2}^{-}\cdots Q_{\cN,p-k+1}^{-}
 Q_{\cN,p-k+1}^{+}Q_{\cN,p-k+1}^{-}\cdots Q_{\cN,p}^{-}
 +Q_{\cN,1}^{+}Q_{\cN,1}^{-}\cdots Q_{\cN,p}^{-}\notag\\
 =C_{p}Q_{\cN,2}^{-}\cdots Q_{\cN,p}^{-}\sPN (H_{p}).
\end{align}
\end{subequations}
The conditions for the second non-linear relation in
Eq.~(\ref{eq:pfsc3}) are apparently given by the `adjoint' of
Eqs.~(\ref{eq:parac}).

An `$\cN$-fold generalization' of quasi-parasupersymmetry proposed
in Ref.~\cite{Ta07a} is also straightforward.

\section{An Example}

Let us now construct a second-order $\cN$-fold
parasupersymmetric quantum system. In the case of $p=2$, the triple
$(\bH,\bQ_{\cN}^{-},\bQ_{\cN}^{+})$ defined in Eqs.~(\ref{eq:pfqs})
and (\ref{eq:pfsc}) is given by
\begin{align}
\bH&=H_{0}(\psi^{-})^{2}(\psi^{+})^{2}
 +H_{1}(\psi^{+}\psi^{-}-(\psi^{+})^{2}(\psi^{-})^{2})
 +H_{2}(\psi^{+})^{2}(\psi^{-})^{2},\\
\bQ_{\cN}^{-}
&=Q_{\cN,1}^{-}(\psi^{-})^{2}\psi^{+}
 +Q_{\cN,2}^{-}\psi^{+}(\psi^{-})^{2},
\label{eq:psc2-}\\
\bQ_{\cN}^{+}
&=Q_{\cN,1}^{+}\psi^{-}(\psi^{+})^{2}
 +Q_{\cN,2}^{+}(\psi^{+})^{2}\psi^{-}.
\label{eq:psc2+}
\end{align}
We recall the fact that the above second-order parasupercharges
(\ref{eq:psc2-}) and (\ref{eq:psc2+}) already satisfy
the nilpotent condition (\ref{eq:pfsc1}) for $p=2$,
$(\bQ_{\cN}^{-})^{3}=(\bQ_{\cN}^{+})^{3}=0$.
From Eqs.~(\ref{eq:inter}) and (\ref{eq:parac}), the
commutativity (\ref{eq:pfsc2}) and the non-linear constraints
(\ref{eq:pfsc3}) for $p=2$
hold if and only if the following conditions
\begin{align}
H_{0}Q_{\cN,1}^{-}=Q_{\cN,1}^{-}H_{1},\qquad
 H_{1}Q_{\cN,2}^{-}=Q_{\cN,2}^{-}H_{2},
\label{eq:2psc1}\\
Q_{\cN,1}^{-}Q_{\cN,2}^{-}Q_{\cN,2}^{+}+Q_{\cN,1}^{-}Q_{\cN,1}^{+}
 Q_{\cN,1}^{-}=C_{2}Q_{\cN,1}^{-}\sPN (H_{1}),
\label{eq:2psc2}\\
Q_{\cN,2}^{-}Q_{\cN,2}^{+}Q_{\cN,2}^{-}+Q_{\cN,1}^{+}Q_{\cN,1}^{-}
 Q_{\cN,2}^{-}=C_{2}Q_{\cN,2}^{-}\sPN (H_{2}),
\label{eq:2psc3}
\end{align}
and their `adjoint' relations
\begin{align}
Q_{\cN,1}^{+}H_{0}=H_{1}Q_{\cN,1}^{+},\qquad
 Q_{\cN,2}^{+}H_{1}=H_{2}Q_{\cN,2}^{+},
\label{eq:2psc1'}\\
Q_{\cN,1}^{+}Q_{\cN,1}^{-}Q_{\cN,1}^{+}+Q_{\cN,2}^{-}Q_{\cN,2}^{+}
 Q_{\cN,1}^{+}=C_{2}\sPN (H_{1})Q_{\cN,1}^{+},
\label{eq:2psc2'}\\
Q_{\cN,2}^{+}Q_{\cN,1}^{+}Q_{\cN,1}^{-}+Q_{\cN,2}^{+}Q_{\cN,2}^{-}
 Q_{\cN,2}^{+}=C_{2}\sPN (H_{2})Q_{\cN,2}^{+},
\label{eq:2psc3'}
\end{align}
are satisfied.
In general, we do not need to solve the `adjoint' conditions.

For the second-order case, we have one new $\cN$-fold
quasi-parasupersymmetry,
namely, that of order $(2,2)$. The conditions are given by
Eqs.~(\ref{eq:2psc1})--(\ref{eq:2psc3'}) but the first-order
intertwining relations (\ref{eq:2psc1}) and (\ref{eq:2psc1'}) are
replaced by the second-order intertwining relations
\begin{align}
H_{0}Q_{\cN,1}^{-}Q_{\cN,2}^{-}=Q_{\cN,1}^{-}Q_{\cN,2}^{-}H_{2},\qquad
 Q_{\cN,2}^{+}Q_{\cN,1}^{+}H_{0}=H_{2}Q_{\cN,2}^{+}Q_{\cN,1}^{+}.
\label{eq:2qpsc1}
\end{align}
Let us next put $C_{2}=2^{\cN+1}$ and
\begin{align}
H_{k}=-\frac{1}{2}\frac{\rmd^{2}}{\rmd q^{2}}+V_{k}(q),
 \qquad Q_{\cN,k}^{+}=\prod_{i=0}^{\cN-1}\left(\frac{\rmd}{\rmd q}
 +W_{k}(q)+\frac{\cN-1-2i}{2}E_{k}(q)\right),
\label{eq:1vrep}
\end{align}
where the product of operators is ordered according to
$\prod_{i=i_{0}}^{i_{1}}A_{i}=A_{i_{1}}A_{i_{1}-1}\cdots A_{i_{0}}$.
Each component $Q_{\cN,k}^{+}$ of $\cN$-fold parasupercharges given in
Eq.~(\ref{eq:1vrep}) is so-called type A $\cN$-fold supercharge, and
the necessary and sufficient condition for two Hamiltonians to be
intertwined by it is already well known~\cite{Ta03a}; the conditions
(\ref{eq:2psc1}) and (\ref{eq:2psc1'}) are satisfied if and only if
\begin{align}
H_{0}&=-\frac{1}{2}\frac{\rmd^{2}}{\rmd q^{2}}+\frac{1}{2}W_{1}(q)^{2}
 +\frac{\cN^{\,2}-1}{24}\left(E_{1}(q)^{2}-2E'_{1}(q)\right)
 -\frac{\cN}{2}W'_{1}(q)-R_{1},
\label{eq:H0}\\
H_{1}&=-\frac{1}{2}\frac{\rmd^{2}}{\rmd q^{2}}+\frac{1}{2}W_{1}(q)^{2}
 +\frac{\cN^{\,2}-1}{24}\left(E_{1}(q)^{2}-2E'_{1}(q)\right)
 +\frac{\cN}{2}W'_{1}(q)-R_{1}\notag\\
&=-\frac{1}{2}\frac{\rmd^{2}}{\rmd q^{2}}+\frac{1}{2}W_{2}(q)^{2}
 +\frac{\cN^{\,2}-1}{24}\left(E_{2}(q)^{2}-2E'_{2}(q)\right)
 -\frac{\cN}{2}W'_{2}(q)-R_{2},
\label{eq:H1}\\
H_{2}&=-\frac{1}{2}\frac{\rmd^{2}}{\rmd q^{2}}+\frac{1}{2}W_{2}(q)^{2}
 +\frac{\cN^{\,2}-1}{24}\left(E_{2}(q)^{2}-2E'_{2}(q)\right)
 +\frac{\cN}{2}W'_{2}(q)-R_{2},
\label{eq:H2}
\end{align}
where $R_{k}$ ($k=1,2$) are constants and the functions $E_{k}$ and
$W_{k}$ ($k=1,2$) must satisfy the following non-linear differential
equations:
\begin{align}
\left(\frac{\rmd}{\rmd q}-E_{k}(q)\right)\frac{\rmd}{\rmd q}
 \left(\frac{\rmd}{\rmd q}+E_{k}(q)\right)W_{k}(q)=0\quad
 \text{for}\quad\cN\geq 2,
\label{eq:condW}\\
\left(\frac{\rmd}{\rmd q}-2E_{k}(q)\right)\left(\frac{\rmd}{\rmd q}
 -E_{k}(q)\right)\frac{\rmd}{\rmd q}\left(\frac{\rmd}{\rmd q}
 +E_{k}(q)\right)E_{k}(q)=0\quad\text{for}\quad\cN\geq 3.
\label{eq:condE}
\end{align}
We note that the formula (\ref{eq:H1}) for $H_{1}$ implies the
following condition among $E_{k}$ and $W_{k}$:
\begin{align}
W_{1}^{2}+\frac{\cN^{\,2}-1}{12}\left(E_{1}^{2}-2E'_{1}\right)
 +\cN W'_{1}-2R_{1}=W_{2}^{2}+\frac{\cN^{\,2}-1}{12}\left(
 E_{2}^{2}-2E'_{2}\right)-\cN W'_{2}-2R_{2}.
\label{eq:EW1EW2}
\end{align}
It is worth pointing out that it is similar to but less restrictive
than the condition for simultaneous type A $\cN$-fold supersymmetry
with two different values of $\cN$, cf. Eqs.~(15) and (16) in
Ref.~\cite{HT06a}.
When the conditions (\ref{eq:H0})--(\ref{eq:condE}) are all satisfied,
it was shown \cite{Ta03a} that the following relations hold
\begin{subequations}
\label{eq:GBDP}
\begin{align}
Q_{\cN,1}^{-}Q_{\cN,1}^{+}=2^{\cN}\pi_{1,\cN}^{[\cN]}(H_{0}),
 \qquad Q_{\cN,1}^{+}Q_{\cN,1}^{-}=2^{\cN}\pi_{1,\cN}^{[\cN]}(H_{1}),\\
Q_{\cN,2}^{-}Q_{\cN,2}^{+}=2^{\cN}\pi_{2,\cN}^{[\cN]}(H_{1}),
 \qquad Q_{\cN,2}^{+}Q_{\cN,2}^{-}=2^{\cN}\pi_{2,\cN}^{[\cN]}(H_{2}),
\end{align}
\end{subequations}
where $\pi_{k,\cN}^{[\cN]}$ are the $\cN$th critical generalized
Bender--Dunne polynomials associated with each system labeled by the
indices $k=1,2$. Substituting Eqs.~(\ref{eq:GBDP}) into the second
condition (\ref{eq:2psc2}), we have
\begin{align}
2^{\cN}Q_{\cN,1}^{-}\pi_{2,\cN}^{[\cN]}(H_{1})+2^{\cN}Q_{\cN,1}^{-}
 \pi_{1,\cN}^{[\cN]}(H_{1})=C_{2}Q_{\cN,1}^{-}\sPN (H_{1}),
\end{align}
and thus obtain a solution to the condition (\ref{eq:2psc2}) as
\begin{align}
\pi_{1,\cN}^{[\cN]}(x)+\pi_{2,\cN}^{[\cN]}(x)=2\sPN (x).
\label{eq:solPN}
\end{align}
Finally, substituting Eqs.~(\ref{eq:GBDP}) and (\ref{eq:solPN}) into
the third condition (\ref{eq:2psc3}), we have
\begin{align}
\pi_{2,\cN}^{[\cN]}(H_{1})Q_{\cN,2}^{-}+\pi_{1,\cN}^{[\cN]}(H_{1})
 Q_{\cN,2}^{-}=2\sPN (H_{1})Q_{\cN,2}^{-}=2Q_{\cN,2}^{-}
 \sPN (H_{2}).
\end{align}
It is evident that this condition is already satisfied since we have
constructed the system so that $H_{1}$ and $H_{2}$ satisfy the
second intertwining relation in Eq.~(\ref{eq:2psc1}). Therefore,
the system (\ref{eq:1vrep})--(\ref{eq:H2}) constitutes a second-order
$\cN$-fold parasupersymmetric quantum system with the monic
$\cN$th-degree polynomial $\sPN$ given by Eq.~(\ref{eq:solPN})
provided that the conditions (\ref{eq:condW})--(\ref{eq:EW1EW2}) are
all satisfied. We note that this system exactly reduces to the
parasupersymmetric quantum system of Rubakov--Spiridonov (RS)
type~\cite{RS88} when $\cN=1$ and $R_{1}+R_{2}=0$.

In our previous paper \cite{Ta07a}, we found that the RS model admits
a generalized $2$-fold superalgebra. In the following, we show
that the above $\cN$-fold parasupersymmetric system also satisfies
a novel non-linear algebra. Using the relation (\ref{eq:GBDP}) and
applying the intertwining relation (\ref{eq:2psc1'}), we obtain for
the system (\ref{eq:1vrep})--(\ref{eq:H2}) the following formulas:
\begin{align}
\bQ_{\cN}^{-}\bQ_{\cN}^{+}&=2^{\cN}\pi_{1,\cN}^{[\cN]}(H_{0})\Pi_{0}
 +2^{\cN}\pi_{2,\cN}^{[\cN]}(H_{1})\Pi_{1},\\
\bQ_{\cN}^{+}\bQ_{\cN}^{-}&=2^{\cN}\pi_{1,\cN}^{[\cN]}(H_{1})\Pi_{1}
 +2^{\cN}\pi_{2,\cN}^{[\cN]}(H_{2})\Pi_{2},\\
(\bQ_{\cN}^{-})^{2}(\bQ_{\cN}^{+})^{2}&=2^{\cN}Q_{\cN,1}^{-}
 \pi_{2,\cN}^{[\cN]}(H_{1})Q_{\cN,1}^{+}\Pi_{0}=2^{\cN}Q_{\cN,1}^{-}
 Q_{\cN,1}^{+}\pi_{2,\cN}^{[\cN]}(H_{0})\Pi_{0}\notag\\
&=2^{2\cN}\pi_{1,\cN}^{[\cN]}(H_{0})\pi_{2,\cN}^{[\cN]}(H_{0})\Pi_{0},\\
(\bQ_{\cN}^{+})^{2}(\bQ_{\cN}^{-})^{2}&=2^{\cN}Q_{\cN,2}^{+}
 \pi_{1,\cN}^{[\cN]}(H_{1})Q_{\cN,2}^{-}\Pi_{2}=2^{\cN}
 \pi_{1,\cN}^{[\cN]}(H_{2})Q_{\cN,2}^{+}Q_{\cN,2}^{-}\Pi_{2}\notag\\
&=2^{2\cN}\pi_{1,\cN}^{[\cN]}(H_{2})\pi_{2,\cN}^{[\cN]}(H_{2})\Pi_{2}.
\end{align}
Hence, we can easily find the following non-linear relation:
\begin{align}
(\bQ_{\cN}^{-})^{2}(\bQ_{\cN}^{+})^{2}+\bQ_{\cN}^{\pm}
 (\bQ_{\cN}^{\mp})^{2}\bQ_{\cN}^{\pm}+(\bQ_{\cN}^{+})^{2}
 (\bQ_{\cN}^{-})^{2}=2^{2\cN}\pi_{1,\cN}^{[\cN]}(\bH)\pi_{2,\cN}^{[\cN]}
 (\bH).
\end{align}
It is interesting to note that this non-linear relation can be
regarded as a generalization of $2\cN$-fold superalgebra. Indeed, if we
restrict the linear space $\fF\times\sV_{2}$ on which the system $\bH$
acts to $\fF\times(\sV_{2}^{(0)}\dotplus\sV_{2}^{(2)})$ (cf. the
definition between Eqs.~(\ref{eq:defpfs}) and (\ref{eq:defpo})),
we have
\begin{align}
\{(\bQ_{\cN}^{-})^{2},(\bQ_{\cN}^{+})^{2}\}=2^{2\cN}\pi_{1,\cN}^{[\cN]}
 (\bH)\pi_{2,\cN}^{[\cN]}(\bH)\bigr|_{\fF\times(\sV_{2}^{(0)}\dotplus
 \sV_{2}^{(2)})}.
\label{eq:2Nalg}
\end{align}
This, together with the trivial (anti-)commutation relations
\begin{align}
\{(\bQ_{\cN}^{-})^{2},(\bQ_{\cN}^{-})^{2}\}=\{(\bQ_{\cN}^{+})^{2},
 (\bQ_{\cN}^{+})^{2}\}=[(\bQ_{\cN}^{\pm})^{2},\bH]=0,
\end{align}
constitutes a type of $2\cN$-fold superalgebra in the sector
$\fF\times(\sV_{2}^{(0)}\dotplus\sV_{2}^{(2)})$. We also note that
the anti-commutation relation (\ref{eq:2Nalg}) is reminiscent of
the one appeared in type C $\cN$-fold supersymmetry, cf.
Eq.~(5.11b) in Ref.~\cite{GT05}. It is not accidental.
Indeed, on one hand it follows from Eqs.~(\ref{eq:2psc1}) and
(\ref{eq:2psc1'}) that $H_{0}$ and $H_{2}$ are intertwined by
$Q_{\cN,2}^{+}Q_{\cN,1}^{+}$, which is the component of
$(\bQ_{\cN}^{+})^{2}$, and on the other hand if we put
\begin{align}
E_{1}=E_{2}=E,\quad W_{1}-\frac{\cN}{2}E_{1}=W,
 \quad W_{2}+\frac{\cN}{2}E_{2}=W+(\cN-\lambda)F,
\end{align}
where $\lambda$ is a parameter, the operator $Q_{\cN,2}^{+}
Q_{\cN,1}^{+}$ is expressed as
\begin{align}
Q_{\cN,2}^{+}Q_{\cN,1}^{+}=&\prod_{i=\cN}^{2\cN-1}\left(
 \frac{\rmd}{\rmd q}+W+(\cN-\lambda)F+\frac{2\cN-1-2i}{2}E\right)
 \notag\\
&\times\prod_{i=0}^{\cN-1}\left(\frac{\rmd}{\rmd q}+W
 +\frac{2\cN-1-2i}{2}E\right),
\end{align}
which is nothing but a type C $2\cN$-fold supercharge with $\cN_{1}=
\cN_{2}=\cN$ (cf. Eq.~(3.22) in Ref.~\cite{GT05}). Hence, $H_{0}$
and $H_{2}$ can be regarded as a type C $2\cN$-fold supersymmetric
pair and thus the formula (\ref{eq:2Nalg}) can be naturally understood.

We also note that as in the case of (ordinary) parasupersymmetry,
quasi-parasupersymmetry of order $(2,2)$ may not produce any new
result over parasupersymmetry of order $2$. The reason is that in
most cases the general solution to the conditions (\ref{eq:2psc2})
and (\ref{eq:2psc2'}) would be given by
\begin{align}
Q_{\cN,1}^{+}Q_{\cN,1}^{-}+Q_{\cN,2}^{-}Q_{\cN,2}^{+}
 =C_{2}\sPN (H_{1}).
\end{align}
In this case, the conditions (\ref{eq:2psc3}) and (\ref{eq:2psc3'})
are equivalent to
\begin{align}
\sPN (H_{1})Q_{\cN,2}^{-}=Q_{\cN,2}^{-}\sPN (H_{2}),\qquad
 Q_{\cN,2}^{+}\sPN (H_{1})=\sPN (H_{2})Q_{\cN,2}^{+},
\end{align}
which is close to the second relation in Eqs.~(\ref{eq:2psc1})
and (\ref{eq:2psc1'}). Hence, in most of the second-order cases, $\cN$-fold
quasi-parasupersymmetry may be identical to $\cN$-fold
parasupersymmetry.

Finally, we would like to refer to the fact that some different
types of generalized supersymmetries have been shown to have intimate
relation with each other. For instance, according to Ref.~\cite{Mo01}
every orthosupersymmetric system~\cite{KMR93b} admits both
parasupersymmetry and fractional supersymmetry~\cite{Du93a}. Combining
this fact with the present results shown in this letter, we conjecture
the existence of `$\cN$-fold generalization' of other supersymmetric
variants such as \emph{$\cN$-fold fractional supersymmetry} characterized
by the following non-linear relation
\begin{align}
(\bQ_{\cN,i})^{p}=C_{p}\sPN (\bH),
\end{align}
\emph{$\cN$-fold orthosupersymmetry} characterized by the following
non-linear relation
\begin{align}
\bQ_{\cN,\alpha}^{\pm}\bQ_{\cN,\beta}^{\pm}=0,\qquad
 \bQ_{\cN,\alpha}^{-}\bQ_{\cN,\beta}^{+}+\delta_{\alpha,\beta}
 \sum_{\gamma=1}^{p}\bQ_{\cN,\gamma}^{+}\bQ_{\cN,\gamma}^{-}
 =C_{p}\delta_{\alpha,\beta}\sPN (\bH),
\end{align}
and so on.

\begin{acknowledgments}
 This work was partially supported by the National Cheng-Kung
 University under the grant No. OUA:95-3-2-071.
\end{acknowledgments}



\bibliography{refsels}
\bibliographystyle{npb}



\end{document}